\documentclass{iopart}%

\usepackage{iopams}

\usepackage{amssymb}
\usepackage{bm}
\usepackage{afterpage}
\usepackage{amsfonts}
\usepackage{amsthm}
\usepackage{graphicx}

\def\beq{\begin{equation}}
\def\eeq{\end{equation}}
\def\nn{\nonumber}
\def\om{\omega}
\def\a{\alpha}
\def\b{\beta}
\def\g{\gamma}
\def\G{\Gamma}
\def\th{\theta}

\def\s{\sigma}
\def\D{\Delta}
\def\d{\delta}

\def\r{{\tilde{\rho}}}
\def\ad{a^\dagger}
\def\rd{{\rm{d}}}
\def\vp{\varphi}
\def\p{\phi}

\def\ra{\rightarrow}
\def\Cc{\mathbb{C}}
\def\Zz{\mathbb{Z}}
\def\Rr{\mathbb{R}}
\def\Zpl{\rm{I\!N}}

\def\Ph{\hat{P}}

\def\H{{\cal H}}
\def\Hp{H_{\rm{2p}}}
\def\dz{\frac{\rd}{\rd z}}
\def\ddz{\frac{\rd^2}{\rd z^2}}
\def\B{{\cal B}}
\def\id{{1\!\!1}}
\def\adq{a^{\dagger 2}}
\def\bz{\bar{z}}
\def\Re{{\rm Re}}
\def\Im{{\rm Im}}
\def\arg{{\rm arg}}
\def\X{\hat{X}}
\def\Y{\hat{Y}}
\def\Z{\hat{Z}}
\def\ch{\textrm{ch}}
\def\sh{\textrm{sh}}
\def\tanh{\textrm{th}}

\def\ba{{\bar{a}}}
\def\Gc{G_{{\scriptscriptstyle ++}}}

\begin{document}


\title{Spectral determinant of the two-photon quantum Rabi model}

 \author{Daniel Braak}
  \address{TP III and
   Center for Electronic Correlations and Magnetism,\\
Institute of Physics,
    University of Augsburg, 86135 Augsburg, Germany}
\ead{daniel.braak@uni-a.de}

  \begin{abstract}
The various generalized spectral determinants ($G$-functions) of the two-photon quantum Rabi model are analyzed with emphasis on the qualitative aspects of the regular spectrum. Whereas all of them yield at least a subset of the exact regular eigenvalues, only the $G$-function proposed by Chen {\it et al.} in 2012 exhibits an explicitly known pole structure which dictates the approach to the collapse point. We derive this function rigorously employing the $\Zz_4$-symmetry of the model and show that its zeros correspond to the complete regular spectrum.  

  \end{abstract}

\maketitle



\section{Introduction}
The quantum Rabi model (QRM) \cite{jaynes_63} is a well-known minimalistic way to describe the interaction of matter (fermions, discrete degrees of freedom) with light (bosons, continuous degrees of freedom). It consists of a spin 1/2 coupled linearly to a single mode of the radiation field,
\beq
H_R=\om\ad a +\D\s_z + g(a+\ad)\s_x. 
\label{Hrabi}
\eeq
The Pauli matrices $\s_{x,z}$ describe the spin 1/2 and $a$ ($\ad$) are the annihilation (creation) operators of the bosonic mode. The Hilbert space is thus
$\H=L^2(\Rr)\otimes\Cc^2$. This model is integrable due to its manifest $\Zz_2$-symmetry  for all  values of frequency $\om$ and coupling $g$ \cite{braak_11}. The small and large coupling regions are also accessible via perturbation theory. For small $g$ the rotating wave approximation is feasible \cite{jaynes_63}, for large $g$ the adiabatic approximation \cite{li_21}. The validity of the adiabatic approximation for large coupling is due to the boundedness of $\id\otimes\s_z$ and the fact that the operator $\ad a +g(a+\ad) +g^2$ is unitarily equivalent to $\ad a$, the Hamiltonian of the uncoupled field mode. Therefore, the qualitative features of the spectral graph as function of $g$ do not change much in going from small to large coupling \cite{rossatto_17}, as long as neither $\om$ nor $\Delta$ become singular \cite{hwang_15,felicetti_20}. Especially the average distance between adjacent levels with the same parity (the eigenvalue of the operator $\Ph_R=-\exp(i\pi\ad a)\otimes\s_z$ generating the $\Zz_2$-symmetry) is always $\omega$, independent of $g$ and $\Delta$ \cite{braak_19}.

There are many generalizations of the QRM which describe various implementations within cavity and circuit QED as well as quantum simulation platforms, e.g. the anisotropic \cite{xie_14} and the asymmetric QRM \cite{wakayama_17}, the Rabi-Stark model \cite{mac_15,eckle_17} and the QRM with non-linear coupling \cite{ng_99}, the so-called two-photon quantum Rabi model (2pQRM) which is the subject of the present investigation. The 2pQRM has received lot of attention recently due to the possibilities to realize it in superconducting circuits \cite{felicetti_18,felicetti_19} or via quantum simulation \cite{felicetti_15}.
After a trivial unitary transformation exchanging $\s_x$ and $\s_z$, the Hamiltonian of the 2pQRM reads
\beq
\Hp = \om\ad a + (a^2 + \adq)\s_z +\D\s_x.
\label{H2p}
\eeq
For convenience, we  measure $\om$ and $\D$ in units of $g$ which is set to 1. 
In contrast to (\ref{Hrabi}), the coupling term in (\ref{H2p}) is not relatively bounded with respect to $\ad a$ \cite{reed_75} and the spectrum undergoes a dramatic change when $\om$ approaches the critical value $\om_c=2$ from above. For $\om$ less than 2, $\Hp$ is no longer bounded from below \cite{ng_99}.
Exactly at $\om=2$, the spectrum contains a discrete and a continuous part while it is pure point for $\om>2$. The change from discrete to continuous spectrum at $\om_c$ has been termed ``spectral collapse'' \cite{felicetti_15}, although this is a misnomer because the discrete spectrum does not collapse to a single point at $\om_c$.
However, exactly this behavior is seen in all numerical evaluations of the spectrum, simply because the Hilbert space used in these calculations has finite dimension. The continuous part of the spectrum at $\om_c$ covers the interval $[E_0,\infty[$ with $E_0=-\om_c/2=-1$ which follows from a direct treatment of this case \cite{lo_20,chan_20}. The critical point $\om_c=2$ allows for a mapping of the eigenvalue problem to a 1D Schr\"odinger equation in an energy-dependent potential. Similar simplifications at criticality appear also in the Rabi-Stark model \cite{mac_15,chen_20}.
    On the other hand, it is not clear from this calculation whether the Hamiltonian exactly at $\om_c$ is the limit of $\Hp(\om)$ for $\om \rightarrow  \om_c^+$ or a singular special case. To answer this question one has to compute the full spectrum for all $\om>\om_c$ and deduce its qualitative behavior in approaching the ``collapse'' point. This can be done without much effort for $\D=0$, where the collapse phenomenon already happens. In this case $\Hp$ is equivalent to a linear combination of generators of $\mathfrak{su}(1,1)$ and can be diagonalized by standard methods \cite{ng_99, penna_18}. The general situation is much more complicated. One reason is the fact that the spectral problem for the 2pQRM and the linear QRM cannot be solved by polynomial ansatz functions which yield f.e. the discrete spectrum of the harmonic oscillator and the hydrogen atom. The simple harmonic oscillator with Hamiltonian
    \beq
    H_{\rm osc}=-\frac{1}{2}\frac{\rd^2}{\rd x^2}+\frac{\om}{2}x^2-\frac{\om}{2},
    \label{hosc}
    \eeq
    has a spectrum of pure point type. The eigenfunctions can be obtained with the ansatz $\psi_n(x)=P_n(x)e^{-\om x^2/2}$, where $P_n(x)$ is a polynomial of degree $n$, $n\in\Zpl_0$. Functions of this type are clearly normalizable in $L^2(\Rr)$, so each $\psi_n(x)$ is an eigenfunction of $H_{\rm osc}$ with eigenvalue $E_n=n\om$. But it is not easy to prove that the discrete spectrum of $H_{\rm osc}$ consists only of the $E_n$, meaning that the set $\{\psi_n\}$ is complete in $L^2(\Rr)$ (supposing in addition the absence of a continuous spectrum). Moreover, the fact that a polynomial ansatz for the eigenfunctions indeed works is due to the simplicity of the differential equation $(H_{\rm osc}-E)\psi=0$ which belongs to the hypergeometric class \cite{slavianov_00}.

    In most cases, the eigenfunctions cannot be obtained by simple ansatz functions and the spectral problem becomes difficult because the normalizability condition in $L^2(\Rr)$,
    \beq
    \langle\psi|\psi\rangle = \int \rd x\ \psi^\ast(x)\psi(x) < \infty,
    \eeq
    for the eigenunction $\psi(x)$ requires knowledge of $\psi(x)$ for all $x$, it is \emph{non-local}. Fortunately, there exists a Hilbert space $\B$, isomorphic to $L^2(\Rr)$, which consists of analytic functions $f(z)$ of the complex variable $z$ \cite{bargmann_61}. In $\B$, the multiplication operator $z$ is adjoint to the derivative $\rd/\rd z$. All elements $f(z)$ in $\B$ satisfy two conditions: The first Bargmann condition means they are holomorphic in all of $\Cc$, which is a local property of $f(z)$. The second Bargmann condition requires finiteness of $\Vert f\Vert$, the Bargmann norm of $f(z)$, obtained form the scalar product,
\beq
\langle f|g\rangle = \frac{1}{\pi}\int \rd z\rd \bz \ e^{-|z|^2}\bar{f}(\bz)g(z).
\label{barg-norm}
\eeq 
This condition is again non-local. But in the case of the harmonic oscillator (and the QRM) the first, local Bargmann condition is sufficient to determine the whole discrete spectrum and prove its completeness. Writing $H_{\rm osc}$ in terms of $a$ and $\ad$, we have
\beq
H_{\rm osc}=\om\ad a = \om z\dz
\eeq
in $\B$, because $a=\dz$ and $\ad=z$. All solutions of $(H_{\rm osc}-E)\psi(z)=0$ read $\psi_E(z)=z^{E/\om}$. They have finite Bargmann norm for all real $E\ge 0$.
So, the second Bargmann condition does not determine the discrete spectrum of $H_{\rm osc}$. But the first Bargmann condition requires $E/\om$ to be a non-negative integer, yielding the spectrum of the harmonic oscillator. Moreover, the set $\{z^n\}$ is clearly a basis for the entire functions in $\Cc$, so it is complete in $\B$. The absence of a continuous spectrum follows at once. Note that the  $\psi_{E_n}(z)$ are not obtained by a special ansatz because the Schr\"odinger equation has no other solutions than $z^{E/\om}$.

We show in the following that also the second Bargmann condition, despite its non-locality, can be used to determine the exact spectrum of Hamiltonians with a single continuous degree of freedom, in this case $\Hp$, by analyzing the singularity structure of the corresponding ordinary differential equation in the complex domain. This is done for all coupling regimes in section \ref{general}. In section \ref{recur}, we discuss generalized spectral determinants proposed in \cite{zhang_17} and \cite{mac_17} for the case $\om>2$. In section \ref{chen} we derive the $G$-function proposed in \cite{chen_12, duan_16}. Conclusions are given in section \ref{concl}.

\section{Singularities of the eigenvalue equation}
\label{general}


We write  an eigenfunction of $\Hp$, $|\psi\rangle\in \B\otimes\Cc^2$ as the column vector $|\psi\rangle=(\p_1(z),\p_2(z))^T$.
The eigenvalue equation $\Hp|\psi\rangle=E|\psi\rangle$ is the following coupled system
\begin{eqnarray}
  \p_1^{(2)} +\om z\p_1^{(1)} +(z^2-E)\p_1 +\D\p_2 &=0,
  \label{orig1}\\
  \p_2^{(2)} -\om z\p_2^{(1)} +(z^2+E)\p_2 -\D\p_1 &=0,
  \label{orig2}
  \end{eqnarray}
where $\p^{(n)}$ denotes the $n$-th derivative of $\p(z)$ with respect to $z$. The system (\ref{orig1},\ref{orig2}) has no singular points for $|z|<\infty$ but a strong irregular singularity at $z=\infty$. Upon elimination of $\p_2$, the equation for $\p_1$ is of fourth order,
\begin{eqnarray}
  \p_1^{(4)} +((2-\om^2)z^2+2\om)\p_1^{(2)} \nn\\
  + (4+2\om E-\om^2)z\p_1^{(1)})
    +(z^4-2\om z^2+2-E+\D^2)\p_1=0.
  \label{orig4}
\end{eqnarray}
The analysis of the irregular singularity at $z=\infty$ according to \cite{ince_12} shows that the singularity is unramified with rank two, or, following the notation of \cite{slavianov_00}, ``$s$-rank'' three. The class of the singularity is four and therefore maximal \cite{ince_12}. This means that for $z\ra \infty$ all solutions are entire and have  the asymptotic form
\beq
\p_1(z)=\exp\left(\frac{\g}{2}z^2+\b z\right)z^\rho\sum_{n=0}^\infty c_nz^{-n},
\label{origasym}
\eeq
where $c_0\neq 0$ and $\g$ is a solution of the equation
\beq
\g^4+(2-\om)\g^2+1=0.
\label{origamma}
\eeq
The four solutions of (\ref{origamma}),
\beq
\g_{1,4}=\frac{1}{2}(-\om\pm\sqrt{\om^2-4}),\qquad
\g_{2,3}=\frac{1}{2}(\om\mp\sqrt{\om^2-4}),
\label{origammas}
\eeq
called characteristic exponents of second kind  with order two, determine the growth type of
the entire function $\p_1(z)$ which also has order two \cite{rubel_96}. The coefficient $\b$ of the linear term in the exponential factor of (\ref{origasym}) is an exponent of second kind with order one \cite{slavianov_00}. 
In general, the formal expansion (\ref{origasym}) of a solution $\p_1(z)$ with a fixed $\g_j$ is only valid in a certain sector $S_l$
(``Stokes sector'') of the complex plane, while in the complementary sectors the asymptotic expansion requires $\g_k\neq\g_j$ \cite{slavianov_00}. The growth type of $\p_1(z)$ is then given by the sector and associated $\g$ with maximal Re$(\g z^2)$.

Now the spectral problem consists in the task to find those solutions of (\ref{orig4}) which have finite Bargmann norm because the first Bargmann condition \cite{bargmann_61} is satisfied by all of them. The opposite case is realized in the linear QRM: The generic solutions are not entire in $\Cc$ but their Bargmann norm is always finite as for the harmonic oscillator \cite{braak_11}.

If $|\p_1|<\infty$, $\p_1(z)$ is an element of $\B$, which is a necessary condition for $|\psi\rangle$ to be an eigenvector of $\Hp$. 
To decide whether $|\p_1|$ is finite, it is sufficient to know the asymptotic behavior of $\p_1(z)$ in each Stokes sector $S_l$,
\beq
\fl
\int_{S_l}\rd z\rd\bz \ e^{-|z|^2}|\p_1(z)|^2 < \infty \Leftrightarrow
\int_{S_l}\rd z\rd\bz \ e^{-|z|^2} \th(|z|-R)|\p_1(z)|^2 <\infty,
\label{asymapprox}
\eeq
where $\th(x)$ denotes the Heaviside step function.
Inequality (\ref{asymapprox}) holds for arbitrary $R<\infty$ because $\p_1(z)$ is entire. Now, for $z\in S_l$ and $|z|>R$,
\beq
\p_1(z)=c_0\exp\left(\frac{\g(S_l)}{2}z^2+\b(S_l) z\right)z^{\rho(S_l)}\big(1+{\cal O}(R^{-1})\big)
\label{asymstokes}
\eeq
for sufficient large but finite $R$
which means that the inequality on the right of (\ref{asymapprox}) will be satisfied if
\beq
\int_{S_l}\rd z\rd\bz \ \th(|z|-R)e^{-|z|^2}\exp\big(\Re(\g(S_l) z^2 +2\b(S_l) z)\big)|z^{\rho(S_l)}|^2 <\infty.
\label{condasym}
\eeq
If (\ref{condasym}) is satisfied for all Stokes sectors $S_l$, $\p_1$ will be normalizable. It is easy to see that the integral in (\ref{condasym}) is finite for $|\g|<1$ in all of $\Cc$ and diverges for $|\g|>1$ in the Stokes sectors with $\Re(\g z^2)>0$, independent of $\beta$ and $\rho$. If $|\g|=1$, the convergence of the integral is determined by $\beta\neq0$. If $\g$ lies on the unit circle and $\beta$ vanishes, (\ref{condasym}) is determined by $\rho$.
From (\ref{origammas}) we can infer three regions in the range of $\om$ with different properties of (\ref{condasym}).
\subsection{$\om>2$}
\label{sec-greater}
In this case, all $\g_j$ are real and non-degenerate. $|\g_{3,4}|>1$ and $|\g_{1,2}|<1$. 
The four Stokes sectors $S_l$ are
\beq
z\in S_l \Leftrightarrow (2l - 1)\pi/4 < {\rm arg}(z) < (2l+1)\pi/4, 
\qquad l=0\ldots 3.
\label{sectors}
\eeq
Solutions with $\g_{1,2}=-\g_{2,1}$ have finite Bargmann norm in all sectors. Solutions with $\g_4<-1$ converge in sectors $S_0$ and $S_2$ and diverge in sectors $S_1$ and $S_3$ while solutions with $\g_3=-\g_4>1$ diverge in sectors $S_0$ and $S_2$ and converge in sectors $S_1$ and $S_3$.
Because the $\g_j$ are non-degenerate, we have $\beta_j=0$ for all $j$, but this has no bearing on the normalizability properties of $\p_1(z)$.

The fact that all asymptotics have $|\g|\neq 1$ entails that in this region the spectrum has a ``pure point'' characteristic \cite{reed_81} and  no continuous part if the continuous spectrum requires asymptotics with $|\g|=1$. That this is likely correct follows from consideration of the known  ``generalized eigenstates'' which are associated with the continuous spectrum \cite{bargmann_61},
\beq
\phi^\vartheta_{x}(z)=\pi^{-1/4}e^{-\frac{x^2}{2}}\exp\left( -e^{2i\vartheta}\frac{z^2}{2} +\sqrt{2}e^{i\vartheta}xz\right),
\label{genstates}
\eeq
with the real phase angle  $0\le\vartheta<2\pi$ and $x\in\Rr$. These states satisfy the orthogonality relation
\beq
\langle \phi^\vartheta_x|\phi^\vartheta_y\rangle=\d(x-y),
\label{ortho}
\eeq
for fixed $\vartheta$, needed to constitute a spectrum
continuous in the parameter $x$. In fact, the image of $\phi_{x_0}^0$ in $L^2(\Rr)$ is $\psi_{x_0}(x)=\d(x-x_0)$, a generalized eigenstate of the position operator $\hat{x}$, while the image of $\p_p^{\pi/2}$ reads $\psi_p(x)=(2\pi)^{-1/2}\exp(ipx)$, a plane wave associated with the eigenbasis of the momentum operator $\hat{p}$. Fourier transformation in $L^2(\Rr)$ corresponds to the unitary transform $z\rightarrow iz$ in $\B$, embedded in the continuous set of isometries of $\B$, $z \rightarrow e^{i\vartheta}z$, each corresponding to a family of generalized eigenfunctions parameterized by the angle $\vartheta$. Each of them is characterized by its exponent of the second kind $\g=-e^{2i\vartheta}$ of order two, the generalized eigenvalue is proportional to the exponent of order one. The position basis ($\vartheta=0$) has $\g =-1$ and the momentum basis ($\vartheta=\pi/2$) $\g=1$.
If the conjecture above is correct, the spectrum of $\Hp$ in the region $\om>2$ has pure point characteristic.
\subsection{$\om<2$}
\label{sec-less}
All $\g_j$ are located on the unit circle, $|\g_j|=1$ and are non-degenerate.
\beq
\g_{1,4}=\frac{1}{2}(-\om\pm i\sqrt{4-\om^2}),\qquad
\g_{2,3}=\frac{1}{2}(\om\mp i\sqrt{4-\om^2}).
\label{imorigammas}
\eeq
Because the $\g_j$ are all different, the $\b_j$ vanish as in case I. As the arguments of the $\g_j$ differ, they do not have common Stokes sectors. The convergence of the integral (\ref{condasym}) in a certain sector $S_l$ of the complex plane where (\ref{origasym}) is asymptotically valid  depends on whether $S_l$ contains the line $\{z|\arg(\pm z)=-\arg(\g)/2\}$. If not, the integral converges.
If the line is contained in $S_l$, the convergence depends on the value of the exponent of the first kind, $\rho(S_l)$. As example, lets consider the sector
$S_\g = \{z| \arg(z)\in[-\arg(\g)/2- \pi/4, -\arg(\g)/2 +\pi/4]\}$.
The integral in (\ref{condasym}) reads then ($\g=e^{i\varphi}$)
\beq
\fl
I(R) = e^{\varphi\Im(\rho)}\int_R^\infty\rd x\int_{-x}^x\rd y\
(x^2+y^2)^{\Re(\rho)}e^{-2\Im(\rho)\arctan(y/x)} e^{-2y^2},
\label{intsg}
\eeq
where the arc from $x-ix$ to $x+ix$ has been replaced by a straight line.
We have
\beq
I(R) < e^{\varphi\Im(\rho)+\mu}2^{\Re(\rho)}\sqrt{\frac{\pi}{2}}\int_R^\infty\rd x \ x^{2\Re(\rho)},
\eeq
with $\mu=\max(0,-\Im(\rho)\pi/2)$.
It follows that (\ref{condasym}) will be satisfied if $\Re(\rho)<-1/2$.
In the present case, $\rho$ is a function of $\g$, $\om$ and $E$,
\beq
  \rho=-\frac{6\g^3+2\om\g^2+(2-\om^2)\g +(4+2\om E-\om^2)\g-2\om}{4\g^3
    +2(2-\om^2)\g}.
  \label{rho}
  \eeq
  With $\g=e^{i\vp}$ we find,
  \beq
  \rho=-\frac{3}{2}-\frac{\om}{2\cos\vp}+\frac{i}{2}\left(\cot\vp+\frac{\om E}{\sin\vp}\right).
  \eeq
This means $\Re(\rho)=-1/2$ for $\g_{1,4}$ and $\Re(\rho)=-5/2$ for $\g_{2,3}$. Normalizable states exist if the exponents $\g_1, \g_4$ are not present in the Stokes sectors containing their critical lines. The other possibility, $\rho=-1/2$ leads to a continuous spectrum, but the generalized eigenstates are not of the form given in (\ref{genstates}), because the exponent $\beta$ vanishes. States with these asymptotics appear in the harmonic oscillator with inverted potential \cite{braak-prep} which is self-adjoint despite the continuous spectrum being unbounded from below, spanning the whole real axis \cite{wienholtz,hellwig,kalf}. As $\Hp$ can be written  in terms of the inverted oscillator and its Fourier transform for $\D=0$, it follows from the Kato-Rellich theorem \cite{reed_75} that $\Hp$ is self-adjoint for $\om <2$, contrary to a conjecture by Ng {\it et al.} \cite{ng_99}. From the two possible values of $\Re(\rho)$, a discrete spectrum cannot be ruled out for $\om <2$. However, its determination is not subject of the present paper.     
  
\subsection{$\om=2$}
\label{sec-critical}
Here, $\g$ takes the values $+1$ and $-1$, both of them doubly degenerate. It follows from (\ref{orig4}) together with (\ref{origasym}) that now $\b$ is determined in terms of $\g$,
\beq
6\b^2\g^2+(2-\om^2)\b^2 +6\g^3+2\om\g^2+(6-2\om^2+2\om E)\g -2\om=0.
\label{beta}
\eeq
For $\g=1$, that means
\beq
\b^2=-(E+1) \quad \Rightarrow \quad \left\{
\begin{array}{lcl}
  E> -1 &  \Rightarrow & \b\in i\Rr,\\
  E<-1 & \Rightarrow & \b \in \Rr,
\end{array}
\right.
\label{gammaplus}
\eeq
and for $\g=-1$,
\beq
\b^2=(E+1) \quad \Rightarrow \quad \left\{
\begin{array}{lcl}
  E> -1 &  \Rightarrow & \b\in \Rr,\\
  E<-1 & \Rightarrow & \b \in i\Rr,
\end{array}
\right.
\label{gammaminus}
\eeq
In both cases, the two-fold degeneracy of $\g$ is lifted by the two possible values for $\b(\g)$, making all four asymptotic solutions linearly independent.
There are now two different parts of the spectrum, $E>-1$ and $E<-1$.
The threshold energy is $E_0=-1$, in accord with \cite{lo_20,chan_20}.

The four Stokes sectors are given in (\ref{sectors}). The ``critical'' line with $\arg(\pm z)=-\arg(\g)/2$ is the real axis for $\g=1$ and the imaginary axis for $\g=-1$. Let's first look at $\g=1$ and $E<-1$. The integral in (\ref{condasym}) reads in sector $S_0$,
\beq
\fl
I_0(R)=\int_R^\infty \rd x\ e^{2\b x}\int_{-x}^x \rd y\ e^{-2y^2}(x^2+y^2)^{\Re(\rho)}e^{-2\Im(\rho)\arctan(y/x)},  
\eeq
as $\b$ is real and will only converge if $\b<0$, independent of $\rho$. In sector $S_1$, we have instead
\beq
\fl
I_1(R)=\int_R^\infty \rd y\ e^{-2y^2}\int_{-y}^y\rd x\ e^{2\b x}(x^2+y^2)^{\Re(\rho)}e^{-2\Im(\rho)[\pi/2-\arctan(x/y)]}
  \eeq
which converges unconditionally. Analogously, the integral converges in sector $S_2$ if $\b > 0$ and independent of $\b$ in $S_3$. For $\g=-1$ and $E<-1$, the same description obtains with the real and imaginary axes interchanged ($\b\in i\Rr$). It follows that the spectral problem is equivalent to a lateral connection problem with non-trivial Stokes multipliers \cite{slavianov_00}, because asymptotic solutions in sectors $S_0$ and $S_2$ ($S_1$ and $S_3$) must have different $\b$ in case $\g=1$ ($\g=-1$) for a state with $E<-1$ to be normalizable. There is seemingly also a second possibility, namely that $\g=1$ in sectors $S_{1,3}$ and $\g=-1$ in sectors $S_{0,2}$ which also corresponds to off-diagonal Stokes matrices but involving the exponents of second kind with order two. We will see later that this case cannot occur, but the first one leads to a normalizable eigenstate if the parameter $E$ is determined such that the lateral connection problem is solved by $\p_1(z)$. Indeed we know from the direct calculation \cite{lo_20,chan_20} (see also \cite{duan_16}) that there are normalizable states below $E<-1$, contrary to the statement made in \cite{mac_17}, where it is claimed that such states cannot exist for any $E$. 

For $E>-1$ and $\g=1$, $\b$ is imaginary and we have in sector $S_0$,
\beq
\fl
I_0(R) = \int_R^\infty\rd x\int_{-x}^x\rd y\
(x^2+y^2)^{\Re(\rho)}e^{-2\Im(\rho)\arctan(y/x)} e^{-2y^2-2\Im(\b)y}.
\eeq
An estimate similar to (\ref{intsg}) yields now that $I_0(R)$ will be finite if $\Re(\rho)<-1/2$. The same condition applies to $S_2$, while the integral converges in sectors $S_{1,3}$. For $g=-1$ (real $\b$) normalizability of a state with $E>-1$ imposes the same condition on $\rho$ and $S_{0,2}$ replaced by $S_{1,3}$.
We find for $\g=1$, $\rho(1)=-8/5$ and for $\g=-1$, $\rho(-1)=0$. According to the argument for the case $\om<2$, one would conclude that there are normalizable solutions with $E>-1$, as the integral in the critical sector converges, provided $\g=1$ determines the asymptotics of $\p_1(z)$. This does not happen because no eigenstate of $\Hp$ at $\om=2$ can have  $\p_1$-asymptotics with $\g=1$. Only $\g=-1$ is a valid exponent of second kind with order two for $\p_1(z)$ satisfying the system (\ref{orig1}),(\ref{orig2}). To see this, let's use the simplification of the system at $\om=2$ mentioned in the introduction, namely
\begin{eqnarray}
  (\partial_z^2+2z\partial_z +z^2)&=(\partial_z+z)^2 -\id&=2\hat{x}^2 -\id,\\
  (\partial_z^2-2z\partial_z +z^2)&=(\partial_z-z)^2 +\id&=-2\hat{p}^2 +\id.
\end{eqnarray}
The system (\ref{orig1}),(\ref{orig2}) becomes
\begin{eqnarray}
  2\hat{x}^2\p_1 &= (E+1)\p_1-\D\p_2,
  \label{phi1eq}\\
  2\hat{p}^2\p_2 &= (E+1)\p_2-\D\p_1.
  \label{phi2eq}
\end{eqnarray}
It can be easily analyzed in the space $L^2(\Rr)$ \cite{lo_20,chan_20}, but the asymptotics of the eigenstates can be inferred already from the special case $\D=0$. We find that $\p_1(z)$ is an eigenstate of the position operator $\hat{x}$ with eigenvalue $x_0=\sqrt{E+1}/\sqrt{2}$ which agrees with the value of $\b$ in (\ref{gammaminus}), the value $\rho=0$ found above for $\g=-1$  and the form of the corresponding generalized eigenstate of $\hat{x}$ in (\ref{genstates}). Therefore, only $\g=-1$ may occur in the asymptotic form of $\p_1$. This is not changed for $\D\neq 0$ because $\p_1$ satisfies then a Schr\"odinger equation on the real line with a potential approaching zero asymptotically \cite{lo_20,chan_20}, therefore the asymptotic form of the scattering states are the same as for $\D=0$. Likewise, for $\p_2$, which is the image of $\p_1$ under Fourier transformation (or $z \rightarrow iz$), its asymptotic form must have $\g=1$.
Because the formal solutions $\p_1(z)$ with $\g=-1$, real $\b$
and $\rho=0$ cannot be made normalizable by adjusting $E$ but are asymptotically generalized eigenstates of $\hat{x}$ for all $E>-1$, they must correspond to the continuous spectrum of  $\Hp$ in this range of energy. Moreover, as $\g$ is fixed to $-1$ for $\p_1(z)$ in each Stokes sector, the option to have normalizable states for $E>-1$, namely $\g=-1$ in sectors $S_{0,2}$ and $\g=1$ in sectors $S_{1,3}$, is ruled out. It follows that no bound states are embedded into the continuum as it appears in the Rabi-Stark model \cite{chen_20}. The same argument applies for $E<-1$, however, here we must assume $\D\neq 0$ to provide a potential allowing for bound states. The asymptotic form $\tilde{\phi}_2(x)$ of a solution to (\ref{phi2eq}) in $L^2(\Rr)$ ($\hat{p}=-i\partial_x$) is obviously $\tilde{\phi}_2(x) \sim \exp(-\sqrt{-(E+1)/2}|x|)$ which corresponds to (\ref{gammaplus}) with two different $\b$ in sectors $S_0$ and $S_2$, the first case mentioned above. The second case, different $\g$ in sectors $S_{0,2}$ and $S_{1.3}$ is not realized. Therefore, the exponent of second kind with order two is unique throughout the complex plane and the Stokes phenomenon manifests only in the subleading exponent of order one. 
\section{Pure point spectrum for $\om>2$}
\label{purepoint}
For $\om>2$,  $\Hp$ has only discrete eigenvalues and normalizable eigenfunctions, because the exponents of second kind with order two are non-degenerate solutions of (\ref{origamma}). The spectral condition amounts to the requirement that the formal solution of (\ref{orig1}),(\ref{orig2}) has only normalizable asymptotics in all four Stokes sectors (\ref{sectors}).
A sufficient condition for this to happen is clearly that only exponents $\g_{1,2}$
appear in the asymptotic expansion of $\p_1(z)$ which occurs only for a discrete set $E_n$, $n=0,1,2 \dots$, of the parameter $E$.

We call any function $G_H(E)$ a ``$G$-function'' for $H$, if it has only zeros on the real axis, coinciding with all (respectively the regular subset of) the discrete eigenvalues $E_n < \infty$ of the self-adjoint operator $H$. $G_H(E)$ is thus a generalized spectral determinant,
\beq
G_H(E)=e^{h(E)}\prod_{n=0}^\infty\left(1-\frac{E}{E_n}\right),
\label{spec-det}
\eeq
where $h(E)$ may be any function bounded from below such that the right hand side of (\ref{spec-det}) converges for all finite $E$. The spectral problem can be considered ``solved'' if one finds  a $G$-function for $H$, which is, of course, not unique. Note that $h(E)$ does not need to be bounded from above -- which will turn out to be rather useful in section \ref{chen}.

It is often convenient to define $G$-functions for different sectors of the spectrum related to invariant subspaces under the symmetries of $H$. In the case of $\Hp$, we have a $\Zz_4$-symmetry generated by $\Ph=\exp(i(\pi/2)\ad a)\s_x$. The four eigenvalues of $\Ph$ are $\pm 1, \pm i$ and label four invariant subspaces of $\cal H$. The $\Zz_2$-subgroup generated by $\Ph^2$ acts only in the bosonic part of $\cal H$ with eigenvalues $\pm 1$ corresponding to even and odd functions of $z$. We denote each invariant subspace as ${\cal H}_{ab}$ with $a,b\in\{+,-\}$ corresponding to the eigenvalues of $\Ph^2$ and $\Ph$: $a= +$ [$-$] for $\p_{1,2}(z)$ even [odd] and $b=\pm$ if $\p_1(iz) =\pm\p_2(z)$ for even $\p_{1,2}(z)$, resp. $\p_1(iz) =\pm i\p_2(z)$ for odd $\p_{1,2}(z)$. These relations between $\p_1(z)$ and $\p_2(z)$ apply to their asymptotic forms (\ref{asymstokes}) and therefore relate the asymptotics of $\p_1(z)$ in Stokes sectors $S_0$ and $S_2$ to the asymptotics of $\p_2(z)$ in sectors $S_1$ and $S_3$ and vice versa. The asymptotic relations among $\Zz_4$-eigenstates will be of importance in the following.

The eigenstates of $\Hp$ belonging to different subspaces may have degenerate eigenvalues at certain points in the parameter space and manifest as level crossings in the spectral graph. In the 2pQRM the level crossings are of two different types: The first, the so-called Juddian case, describes level crossings between ${\cal H}_{++}$ and  ${\cal H}_{+-}$, respectively between
${\cal H}_{-+}$ and  ${\cal H}_{--}$. It has been investigated in \cite{emary_02,zhang_13}. The two-fold degeneracy of the Juddian solutions corresponds to a two-dimensional representation of $\Zz_4/\Zz_2 \sim \Zz_2$. The second type describes also two-fold degeneracies but between subspaces with different eigenvalues of $\Ph^2$. They have been studied in \cite{mac_19}. In both cases the eigenfunctions can be given in terms of either elementary functions or functions belonging to the hypergeometric class. The degenerate eigenvalues belong to the exceptional spectrum \cite{braak_11} and are not the subject of the present paper which concerns the regular spectrum comprising the non-degenerate eigenvalues. It has to be noted that the non-degenerate spectrum may contain certain eigenvalues at isolated points in parameter space (the so-called non-degenerate exceptional spectrum) which have to be dealt with separately, see \cite{braak_19} and section \ref{chen}. We confine the following discussion to the subspace $\H_{++}$, containing only even functions of $z$.
\subsection{The $G$-functions based on recurrence relations}
\label{recur}
The eigenvalue $+1$ of $\Zz_4/\Zz_2$ means that a solution of (\ref{orig1}),(\ref{orig2}) belonging to $\H_{++}$ satisfies $\p_1(iz)=\p_2(z)$. If $\p_{1,2}(z)$ are even, their expansion in powers of $z$ around $z=0$ reads
\beq
\p_1(z)=\sum_{n=0}^\infty a_nz^{2n}, \quad \p_2(z)=\sum_{n=0}^\infty (-1)^na_nz^{2n},
\label{z4rel}
\eeq
with arbitrary $a_0\neq 0$.
From (\ref{orig1}) we deduce the following three-term recurrence relation for the
$a_n$,
\beq
(2n+2)(2n+1)a_{n+1}+(2n\om-E+(-1)^n\D)a_n+a_{n-1}=0,
\label{origrecur}
\eeq
and $a_n=0$ for $n<0$.
This recurrence relation is the starting point for the $G$-functions proposed in \cite{zhang_17} and \cite{mac_17}.
A Poincar\'e analysis of (\ref{origrecur})
with the asymptotic ansatz
\beq
a_n \sim \frac{x^n}{n!} \quad \textrm{for} \quad n\ra \infty,
\label{asympa}
\eeq
leads to the equation for $x$,
\beq
x^2+\frac{\om}{2}x+\frac{1}{4}=0,
\label{Z-equ}
\eeq
with solutions
\beq
\fl
x_+=\frac{1}{4}\left(\sqrt{\om^2-4}-\om\right) = \frac{\g_1}{2}, \quad
x_-=-\frac{1}{4}\left(\sqrt{\om^2-4}+\om\right) = \frac{\g_4}{2}.
\label{asympx}
\eeq
As $|x_-|>|x_+|$, the Perron-Kreuser theorem \cite{gautschi_67} states that there exist a minimal solution to the recurrence (\ref{origrecur}) with
\beq
\frac{a_{n+1}}{a_n} \sim \frac{x_+}{n} \quad \textrm{for} \quad n\ra \infty,
\eeq
and (\ref{asympa}) suggests that the asymptotic form of $\p_1(z)$ for $|z|\ra\infty$ reads in this case
\beq
\p_1(z) = \exp\left(\frac{\g_1}{2}z^2 +r(z)\right)\left[c_0+ \textrm{less divergent terms}\right],
\label{asympZ}
\eeq
where $r(z)$ grows at most linear in $z$ but is undetermined by the asymptotics of $a_n$.
A function with these leading asymptotics in all Stokes sectors corresponds to a normalizable solution of (\ref{orig4}). Because Pincherle's theorem   states that the minimal solution of a three-term recurrence relation, if it exists, is determined by a convergent continued fraction \cite{gautschi_67} (see also \cite{braak_cont}), one may write down a $G$-function for the spectral problem in $\H_{++}$ as follows \cite{zhang_17},
\beq
G_Z(E)=\frac{E-\D}{2} -V_1(E),
\label{zhangG}
\eeq
with
\beq
\fl
V_n(E)= \frac{a_n^{{\textrm{\tiny min}}}}{a_{n-1}^{{\textrm{\tiny min}}}}=\frac{-1}{2n\om +(-1)\D-E+(2n+2)(2n+1)V_{n+1}}, \quad n=1,2,\ldots,
\eeq
and $(E-\D)/2=a_1/a_0$, as determined from the initial condition for (\ref{origrecur}). The vanishing of $G_Z(E)$ at some point $E_s$ means that the minimal solution of (\ref{origrecur}) whose first terms are given by the
continued fraction $V_1(E_s)=a_1^{{\textrm{\tiny min}}}/a_{0}^{{\textrm{\tiny min}}}$ matches the initial condition $a_n=0$ for negative $n$, which is dictated by the analyticity of $\p_1(z)$ at $z=0$. The state is therefore normalizable because the asymptotic behavior of $\p_1(z)$ apparently contains only the exponent $\g_1$, as guaranteed by the minimality of the solution to (\ref{origrecur}). It follows that $E_s$ belongs to the pure point spectrum of $\Hp$ if $G_Z(E_s)=0$.

Zhang's approach is certainly the simplest and most elegant way to obtain a $G$-function for $\Hp$ in the regime $\om>2$. Following Okubo \cite{okubo_63},  Maciejewski and Stachowiak \cite{mac_17} have proposed a method which is based as well on the determination of those solutions of a three-term (matrix) recurrence relation which correspond to normalizable asymptotics for $\p_{1,2}$ because they contain only exponents $\g_{1,2}$. After ``gluing'' them to the initial conditions as above, they obtain another $G$-function ((43) in \cite{mac_17}) which is given in terms of a contour integral over the solution of an auxiliary differential equation.

Coming back to (\ref{Z-equ}), we have $|x_+|=|x_-|$ for $\om\le 2$, a minimal solution does not exist and the continued fraction does not converge. One would thus conclude with the authors of \cite{mac_17} that in this regime there are no normalizable solutions of (\ref{orig1}),(\ref{orig2}) satisfying in addition $\p_1(iz)=\p_2(z)$. However, as we have seen in section \ref{general}, the pure point spectrum is not empty for $\om=2$ and normalizable states could exist also for $\om <2$. The discrepancy is resolved by noting that the Poincar\'e analysis of the asymptotic behavior of the recurrence relation only yields the leading, most divergent term in the asymptotic expansion and says nothing about the subleading terms which determine normalizability in the case $|\g|=1$. The formal solution (\ref{z4rel}) may or may not be normalizable even if no minimal solution of (\ref{origrecur}) exists.
Another consequence is the missing Stokes phenomenon: The minimal solution of the recurrence relation for $\p_1(z)$ fixes the allowed exponent to $\g_1$, independent of the argument of $z$. $\p_1(z)$ should approach zero in sectors $S_{0,2}$ whereas $\p_2(z)$ diverges there, obviously contradicting (\ref{orig1}) if $\D\neq 0$. In fact, the form (\ref{asympZ}) of the asymptotic behavior of $\p_1(z)$ is only valid in sectors $S_{1,3}$ where the term $\exp(\g_1z^2/2)$ diverges for $|z|\ra\infty$. In the sectors $S_{0,2}$, where it is recessive, subdominant contributions in the recurrence (\ref{origrecur}) determine the asymptotic behavior of $\p_1(z)$, given in these sectors as $\p_1(z)\sim \exp(\g_2z^2/2)$. 
Because the asymptotics of the recurrence relation does not provide the correct behavior of $\p_1(z)$ in all Stokes sectors, one may ask whether indeed all normalizable states satisfying (\ref{orig1}), (\ref{orig2}) are related to the minimal solution of (\ref{origrecur}). For certain values of $E$, there could be an entire function $\p(z)$ behaving asymptotically like $\exp(\g_4z^2/2)$ in sectors $S_{0,2}$ and as $\exp(\g_3z^2/2)$ in $S_{1,3}$. This function would be normalizable although it belongs to a dominant solution of (\ref{origrecur}). If this happens, there would be elements of the spectrum not given by zeros of $G_Z(E)$. The same argument applies to the $G$-function of \cite{mac_17}. The phenomenon occurs (for different reasons) in the QRM with linear coupling, where the non-degenerate exceptional spectrum cannot be obtained by Schweber's $G$-function based on  continued fractions \cite{braak_16}.

In the present case of the 2pQRM, such an exceptional eigenstate cannot occur, at least not in the non-degenerate part of the spectrum, due to the relation $\p_1(z)=\a\p_2(iz)$ with $\a\in\{\pm1,\pm i\}$. It entails that if $\p_1(z)$ has strict recessive behavior in a given sector with exponent $\g$, $\p_2(z)$ will have exponent $-\g$ in the same sector and would not be normalizable if $\g$ is either $\g_3$ or $\g_4$. Indeed, the $\Zz_4$-symmetry of the solution shows that exponents $\g_1$ and $\g_2$ appear in the asymptotics of both $\p_1$ and $\p_2$ in all sectors but not $\g_3$ or $\g_4$ if the state is normalizable.
Thus we may conclude that  $G_Z(E)$ yields the complete spectrum for $\om >2$. The argument is less clear for the $G$-function based on Okubo's method \cite{mac_17}. Because the factorial power series in $x_+$  is actually a double series (see (34) in \cite{mac_17}, $x_+$ is called $u_0$), it is not obvious that the asymptotics have the form $\p_1(z)\sim \exp(\pm x_+z^2)$ in all sectors.

\begin{figure}
  \includegraphics[width=0.6\linewidth]{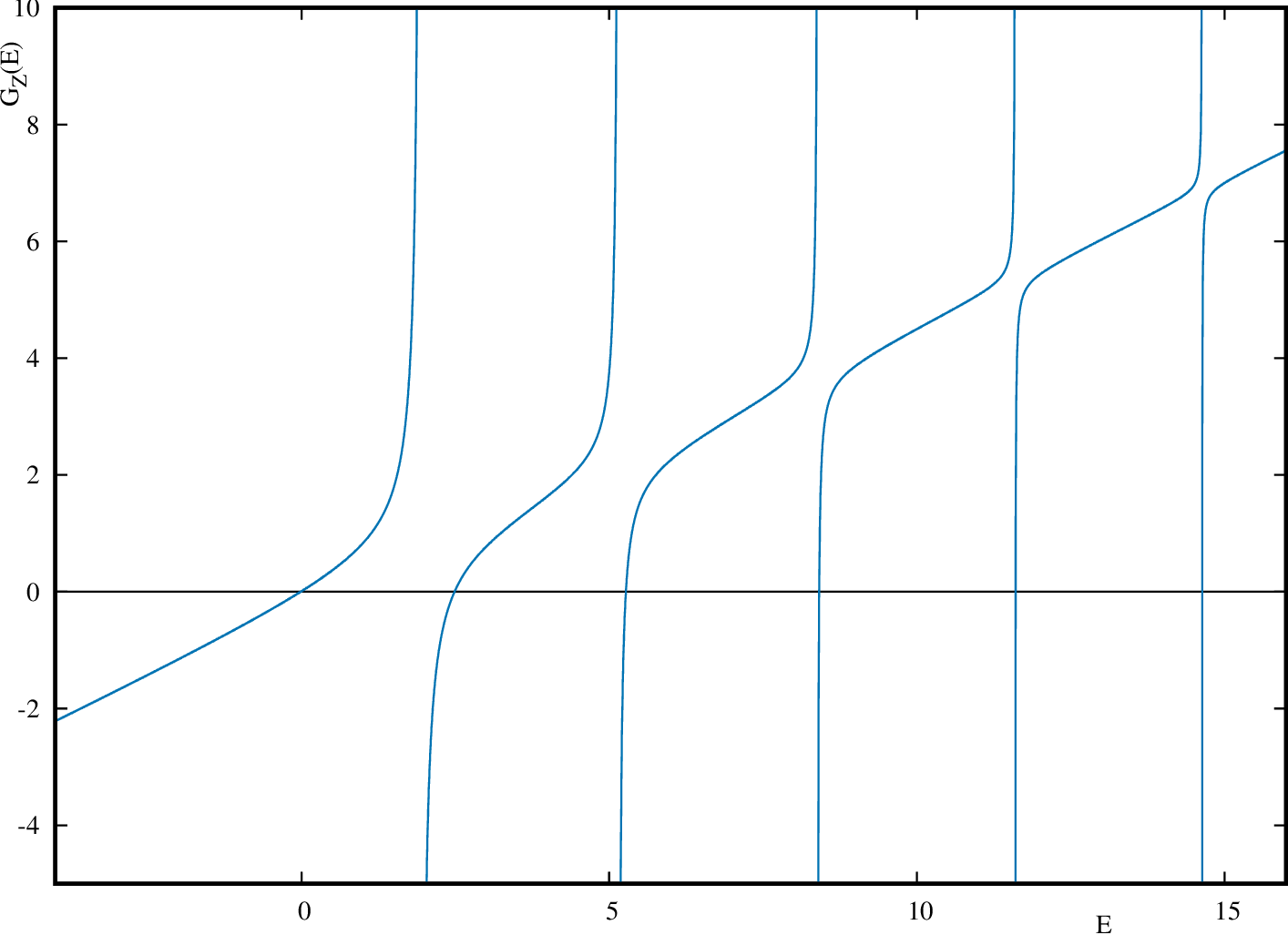}
  \caption{Zhang's $G$-function in sector $\H_{++}$ for parameters $\om=2.5$ and $\D=0.7$. The poles of $G_Z(E)$ move closer to its zeros $E_n$ for larger $E$.}
  \label{zgfunction}
\end{figure}
Zhang's $G$-function is shown in Figure \ref{zgfunction}. It has poles and zeros
which move closer together for higher values of $E$, as typical for a continued fraction. The resolution of higher zeros poses thus known numerical problems but the major drawback of $G_Z(E)$ is the lack of qualitative knowledge about the distribution of its zeros, constituting the regular spectrum of $\Hp$ in $\H_{++}$. The poles are all of first order and $G_Z(E)$ is monotonous, but the location of the poles is just as unknown as the zeros -- they do not allow to restrict the distribution of the zeros as the $G$-functions for the QRM which possess poles at known positions, namely $E=n\om$, with $n$ a non-negative integer \cite{braak_11,zhong_13,mac_14}. The $G$-function proposed in \cite{mac_17} has the same problem: The exact eigenvalues of $\Hp$ can be extracted by a numerical procedure but this has no advantage compared with direct diagonalization in a truncated Hilbert space because the graph of the $G$-function has no properties, say the position of maxima or minima, which are known in closed form (see Figures 1 and 2 in \cite{mac_17}). On the other hand, exact diagonalization in a state space of finite dimension suggests the collapse of the complete spectrum to a single point at $\om_c$, a numerical artifact of the truncation. But the $G$-functions based only on the recurrence relation (\ref{origrecur}) offer no option to study the pure point spectrum arbitrarily close to $\om_c$ in a qualitative way.
\subsection{The $G$-functions based on the $\Zz_4$-symmetry}
\label{chen}
The recurrence (\ref{origrecur}) implements the $\Zz_4$-symmetry directly via the relation (\ref{z4rel}) between the series expansions of $\p_1(z)$ and $\p_2(z)$. It is therefore automatically satisfied by all solutions of (\ref{origrecur}), independent of their normalizability. The question arises whether the $\Zz_4$-symmetry can be used to discern normalizable from non-normalizable solutions as in the linear QRM. In the latter case, the two regular singular points of the eigenvalue equation are located at $z=\pm g$ and mapped onto each other by the $\Zz_2$-symmetry $z\ra -z$ of that model. This can be used to construct a $G$-function which has only zeros at those energies corresponding to solutions $\p_1(z)$,$\p_2(z)$ which are analytic in the whole complex plane by using Frobenius expansions which are analytic in the vicinity of only one of the singular points -- the symmetry requirement $\p_2(z)=\p_1(-z)$ is then equivalent to analyticity at the other singular point \cite{braak_11}.

The situation is different for the 2pQRM: Here there is only a single singular point at $z=\infty$ but with higher rank, preventing all formal solutions to be normalizable, although all are entire. The invariant points of the map $z\ra iz$ are $z=0,\infty$ as in the QRM and $z=0$ is an ordinary point of the system (\ref{orig1}),(\ref{orig2}). The symmetry generator acts trivially on the point $z=\infty$ and maps the asymptotics of $\p_{1,2}(z)$ onto each other: $\g_1\leftrightarrow\g_2$ and $\g_3\leftrightarrow\g_4$. Clearly, because $z\ra iz$ is an isometry of $\B$, the normalizable forms are mapped onto themselves, as well as the non-normalizable ones.

\begin{figure}
  \includegraphics[width=0.6\linewidth]{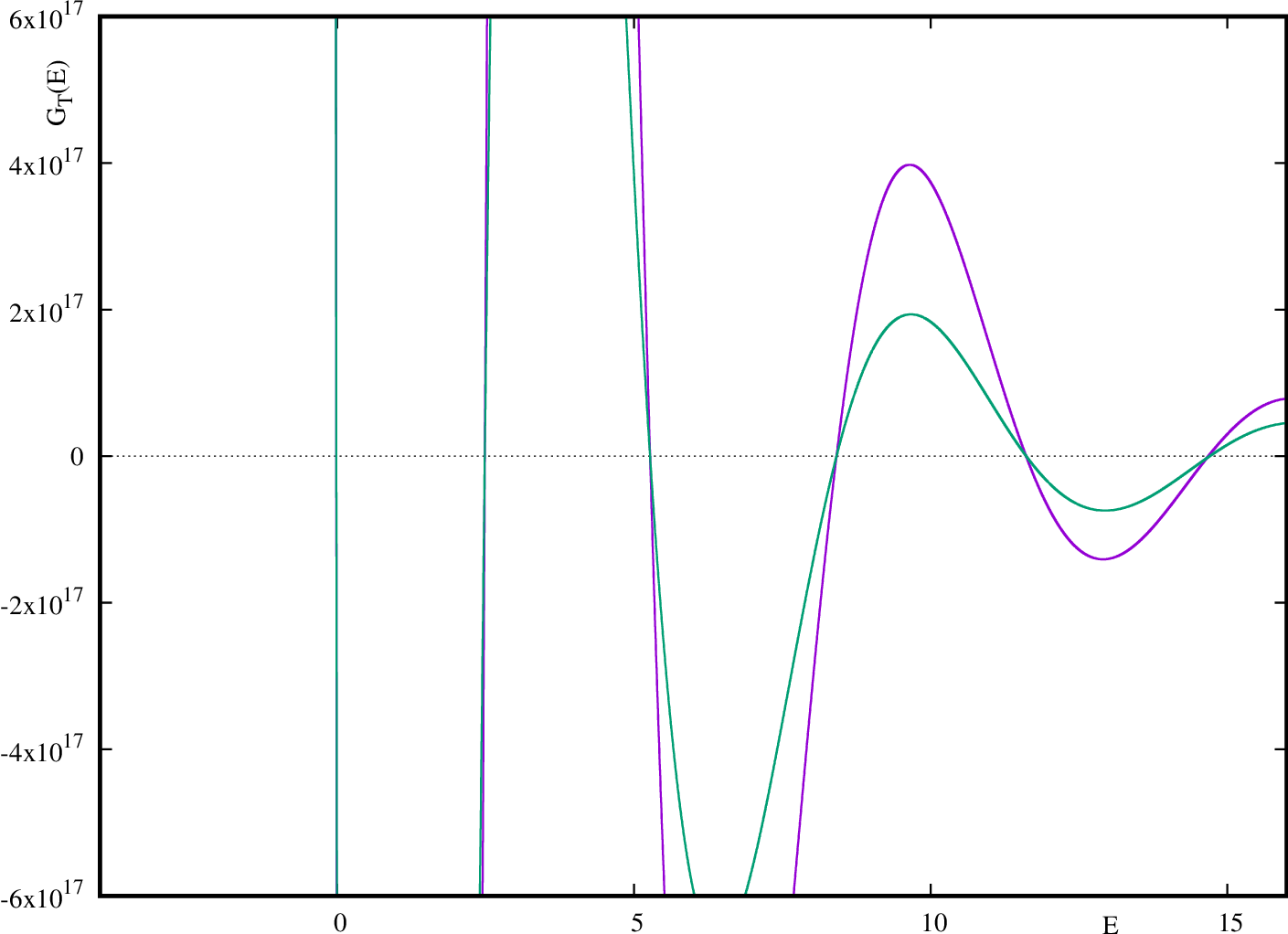}
  \caption{Trav{\v e}nec's $G$-function for $\om=2.5$ and $\D=0.7$ with $z_0=5+5i$. The real (magenta) and imaginary (green) parts of $G_T(E)$ share the same zeros which are located at the eigenenergies of $\Hp$.}
  \label{tgfunction}
\end{figure}

Trav{\v e}nec \cite{travenec_12} has attempted to use the $\Zz_4$-symmetry to derive a $G$-function for $\Hp$ in analogy with the linear QRM as follows. First, separate the asymptotic factor $\exp(\g_1z^2/2)$ from the expansions of $\p_{1,2}$,
\beq
\p_j(z)=e^{\frac{\g_1}{2}z^2}\bar{\p}_j(z), \quad
\bar{\p}_j(z)=\sum_{n=0}^\infty a^j_n z^{2n}, \quad j=1,2.
\eeq
The coefficients $a^j_n$ satisfy the coupled recurrence relations
\begin{eqnarray}
  \fl
(2n+2)(2n+1)a^1_{n+1} &=& (E-\g_1-2n(2\g_1+\om))a_n^1 -\D a_n^2,
  \label{trav1}\\
  \fl
(2n+2)(2n+1)a^2_{n+1} &=& (-E-\g_1-2n(2\g_1-\om))a_n^2 + \D a_n^1 +2\om\g_1a^2_{n-1},
\label{trav2}
\end{eqnarray}
with initial condition $a_0^1=a_0^2=1$.
The symmetry $\p_1(iz)=\p_2(z)$ is not assumed beforehand in the system (\ref{trav1}),(\ref{trav2}).    
The $G$-function is then defined as
\beq
G_T(E) = \p_2(-iz_0)-\p_1(z_0),
\label{Gtrav}
\eeq
with $z_0\neq 0$  an arbitrary point in the complex plane. The zeros of $G_T(E)$ should not depend on the choice of $z_0$. The problem with this approach is that the implementation of the symmetry condition in (\ref{Gtrav}) cannot discern between normalizable and non-normalizable solutions. (Non-)normalizability is an invariant property of any function under $z\ra iz$, as we see from the behavior of the possible asymptotic forms.
Moreover, the initial condition $\p_1(0)=\p_2(0)$ enforces the symmetry for any solution of (\ref{trav1}),(\ref{trav2}). The function $G_T(E)$ vanishes therefore identically \cite{mac_15-2}. Nevertheless, a straightforward numerical implementation of (\ref{Gtrav}) yields a function which does not vanish identically and possesses zeros at the exact eigenenergies of $\Hp$ in $\H_{++}$. Figure~\ref{tgfunction} shows a plot of $G_T(E)$ for the same parameters as in Figure~\ref{zgfunction}. The zeros of $G_Z(E)$ and $G_T(E)$ coincide, but the non-zero values of $G_T(E)$ are extremely large.
Both observations, the numerical validity of $G_T(E)$ and its large values, have the same origin, namely the fact that a three-term recurrence relation has a unique minimal solution. The numerical evaluation of (\ref{trav1}),(\ref{trav2}) will give valid results only if the initial conditions including the parameter $E$ are fine-tuned to yield the minimal solution \cite{gautschi_67, braak_13}. The evaluation of dominant solutions is always numerically unstable, leading to the exponential accumulation of rounding errors which produce the huge non-zero values of $G_T(E)$ for any value of $E$ not belonging to the spectrum. Only if $E$ coincides with some $E_n$, the recurrence yields the (normalizable) minimal solution and the numerics will reproduce $G_T(E)=0$, the value of $G_T(E)$ for \emph{any} $E$ if evaluated exactly. The same mechanism appears also in the $G$-function for the oscillator with quartic anharmonicity proposed by Lay \cite{lay_97}. 
Trav{\v e}nec's $G$-function is thus numerically exact in the same sense as Lay's method to compute the spectrum of the quartic oscillator. But as it is only non-zero due to numerical instabilities, $G_T(E)$ cannot be used to draw qualitative inferences about the spectrum in an analytical way because it has no exactly known properties. In this respect, there is no difference between Trav{\v e}nec's \cite{travenec_12}, Zhang's \cite{zhang_17} or Maciejewski's $G$-functions \cite{mac_17}.

The root of the problem is the invariance of the singularity of (\ref{orig1}),(\ref{orig2}) and its asymptotic solutions under the symmetry operation $z\ra iz$. Therefore, this system of differential equations has to be transformed in such a way that the symmetry
acts non-trivially on its singularity structure.

\begin{figure}
  \includegraphics[width=0.6\linewidth]{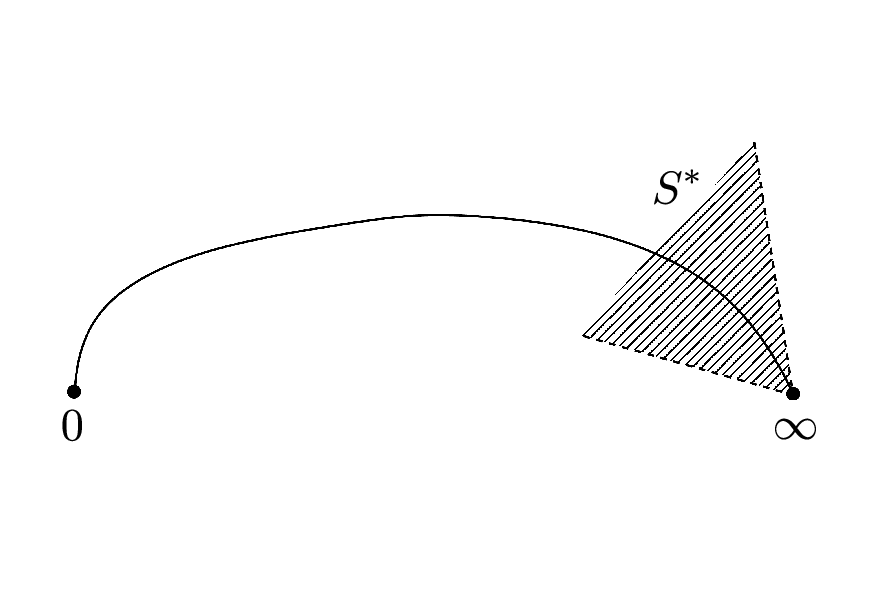}
  \caption{The global behavior of the third solution of (\ref{eq3}) for $E_1/\om_1\notin\Zz$. The branch-cut runs from zero to $\infty$ within the sector $S^\ast$.}
  \label{branch}
\end{figure}

The bosonic Bogoliubov transformation is well established in quantum optics to describe squeezed light \cite{klimov_09} and was used by Chen {\it et al.} in \cite{chen_12} to derive a $G$-function for the 2pQRM. The derivation was heuristic in the sense that it did not establish a proof that the zeros of this function correspond to normalizable eigenstates of $\Hp$. We shall show that this is indeed the case. In $L^2(\Rr)$, the Bogoliubov transformation amounts to the isometry
\beq
I_\th[\psi](x)=e^{\th/2}\psi(e^\th x) =
\exp\left(\frac{\th}{2}\left(a^2-\adq\right)\right)\psi(x),
\eeq
for $\th\in\Rr$. Because $\th$ is real, the operator acting on $\psi(x)$ is a unitary representation of an element of $SU(1,1)$ with generator
\beq
\frac{1}{2}\left(a^2-\adq\right)= \Y-\X,
\eeq
where $\X=(1/2)\adq$, $\Y=(1/2)a^2$ and $\Z=\ad a +(1/2)\id$ fulfill the commutation relations of $\mathfrak{sl}_2(\Rr)$,
\beq
    [\Z,\X]=2\X, \quad [\Z,\Y]=-2\Y, \quad [\X,\Y]=-\Z.
\eeq
The creation and annihilation operators transform under $I_\th$ as
\begin{eqnarray}
  a_\th &= I_\th a I^{-1}_\th &=\ch(\th) a +\sh(\th)\ad,\nn\\
  \ad_\th &= I_\th \ad I^{-1}_\th &=\ch(\th)\ad +\sh(\th) a.\nn
  \end{eqnarray}
Applying $I_\th$ in the Bargmann space $\B$ to the system (\ref{orig1}),(\ref{orig2}) leads to
\begin{eqnarray}
   \om_1z\vp_1^{(1)} - E_1\vp_1 +\D\vp_2 &=& 0,
  \label{eq1}\\
  \G\vp_2^{(2)} + \om_2 z\vp_2^{(1)} +(\G z^2+E_2)\vp_2 - \D\vp_1 &=& 0,
  \label{eq2}
\end{eqnarray}
if $\tanh(2\th)=-2/\om$ which requires $\om>2$.
We have set $\vp_j(z)=I_\th[\p_j](z)$.
The parameters are
\begin{eqnarray}
  \G &=& 2\ch(2\th),\nn\\
  E_1 &=& E - \sh(2\th) -\om\sh^2(\th),\nn\\
  E_2 &=& E + \sh(2\th) -\om\sh^2(\th),
  \label{params}\\
  \om_1 &=& -\sh(2\th)\left(\frac{\om^2}{2}-2\right) > 0,\nn\\
  \om_2 &=&  \sh(2\th)\left(\frac{\om^2}{2}+2\right) < 0\nn.
\end{eqnarray}
The transformed system (\ref{eq1}),(\ref{eq2}) has a singularity structure different from (\ref{orig1}),(\ref{orig2}) but it is still invariant under $z\ra -z$, so the solutions are even or odd functions of $z$. Upon elimination of $\vp_2(z)$, we obtain an equation of third order for $\vp_1(z)$,
\beq
\fl
\vp_1^{(3)}+\left(b_1z+\frac{b_2}{z}\right)\vp_1^{(2)}
+(z^2+b_3)\vp_1^{(1)} +\left(-\frac{E_1}{\om_1}z+\frac{b_4}{z}\right)\vp_1=0,
\label{eq3}
\eeq
with abbreviations
\begin{eqnarray}
  b_1 &=& \frac{\om_2}{\G},\nn\\
  b_2 &=& 2-\frac{E_1}{\om_1},
  \label{abbrev}\\
  b_3 &=& \frac{E_2\om_1 - E_1\om_2 +\om_1\om_2}{\G\om_1},\nn\\
  b_4 &=& \frac{\D^2-E_1E_2}{\G\om_1}\nn.
\end{eqnarray}
The equation (\ref{eq3}) has a regular singular point at $z=0$ and an unramified irregular singular point at $z=\infty$ of rank two and class two. The indicial equation at $z=0$ has three solutions corresponding to exponents $\rho=0,1, E_1/\om_1$ in the Frobenius expansion
\beq
\vp_1(z)=z^\rho\sum_{n=0}^\infty a_nz^n,
\eeq
around $z=0$. The first two correspond to even and odd solutions analytic at $z=0$. The third solution with $\rho=E_1/\om_1$ is only entire if $E_1/\om_1$ is a non-negative integer. This is in close analogy to the linear QRM, where the exceptional spectrum is characterized by $E/\om\in \Zpl_0$ \cite{braak_19}.
For now we assume $E_1/\om_1\notin \Zz$. The other cases will be discussed later. The irregular singular point is $z=\infty$, as in (\ref{orig4}). Rank two entails that the asymptotics of $\vp_1(z)$ have again the form (\ref{origasym}). The equation for $\g$ reads now
\beq
\g^3+\frac{\om_2}{\G}\g^2 +\g=0,
\label{gamma}
\eeq
with solutions $0$, $\om/2$ and $2/\om$. The first solution $\g=0$ corresponds to the fact that the class of the singularity is not maximal \cite{ince_12}. The indicial equation has degree one, meaning that there is a solution $\vp_1(z)$ where $z=\infty$ could be a regular singular point. The asymptotic expansion has then the form
\beq
\vp_1(z)=z^\r \sum_{n=0}^\infty c_nz^{-n}.
\label{regasym}
\eeq
Plugging this expansion into (\ref{eq3}), we find that all odd coefficients vanish, $c_{2m+1}=0$, for $m\in\Zpl_0$, and the exponent of first kind $\r$ has the unique value
$\r=E_1/\om_1$. The even coefficients $c_{2m}$ are determined recursively
\begin{eqnarray}
  c_n&=\frac{1}{n}\Big(\big[b_1(\r-n+1)(\r-n+2)+b_3(\r-n+2)+b_4\big]c_{n-2}
  \label{recur-c}\\
  &+(4-n)(\r-n+3)(\r-n+4)c_{n-4}\Big),\nn
\end{eqnarray}
for $n\ge 2$ and $c_{-2}=0$. The coefficients in the three-term recurrence relation (\ref{recur-c}) are diverging for $n\ra\infty$, therefore the convergence radius of (\ref{regasym}) is zero. There is, as expected, no solution of (\ref{eq3}) regular at $z=\infty$ if $\r\notin\Zz$. Nevertheless, there is a certain Stokes sector $S^\ast$ where (\ref{regasym}) is asymptotically valid for $z\in S^\ast$ and $|z|\ra\infty$. This means that in $S^\ast$ the branching behavior of $\vp_1(z)$ is correctly described by the exponent $\r\notin \Zz$ which is the \emph{same} as the non-integer exponent $\rho=E_1/\om_1$ of the third Frobenius solution in the vicinity of $z=0$. The global behavior of this solution can be described by a branch-cut running from $z=0$ to $z=\infty$ as shown in Figure~\ref{branch}.
The two other solutions have non-zero exponents of second kind $\g$ and $\b=0$. One of them ($\g=\om/2$) has not normalizable asymptotics and the other ($\g=2/\om$) is normalizable if it is asymptotically valid in $S_{0,2}$. For generic values of $E$, both solutions analytic at $z=0$ will exhibit the exponent $\om/2$ in sectors $S_{0,2}$ and are thus not elements of $\B$. If $E$ belongs to the spectrum of $\Hp$ in $\H_{++}$, the even solution of (\ref{eq3}) will be entire and normalizable with asymptotics described by $\g=2/\om$ in $S_{0,2}$. The asymptotic exponent $\r$ for $\g\neq 0$ depends on $E$ and $\g$ but has no influence on the normalizability of $\vp_1$. Because $\g\neq0$, its behavior close to $z=\infty$ is determined by the essential singularity of the factor $\exp((\g/2)z^2)$ which is isolated if $\vp_1$ is entire. 

Up to now, we have not employed the $\Zz_4$-symmetry which is implemented in $\B$ by the operator $T$ as follows
\beq
T[\psi](z)=\psi(iz)=\exp\left(i\frac{\pi}{2}z\dz\right)\psi(z)=
e^{-i\pi/4}\exp\left(i\frac{\pi}{2}\Z\right)\psi(z).
  \label{origt}
  \eeq
  $T$ is therefore an element of $SL_2(\Cc)$ like the Bogoliubov transformation $I_\th$. To compute $T_\th=I_\th TI_\th^{-1}$ and to find the normal ordered form
  \beq
  g = \exp(\a_x\X)\exp(\a_z\Z)\exp(\a_y\Y), \quad \a_{x,y,z}\in\Cc,
  \eeq
  of any group element $g$ in $SL_2(\Cc)$, it is convenient to use the two-dimensional representation with
  \beq
  \Z=\left(\!\!
  \begin{array}{cc}
    1&0\\
  0&-1
  \end{array}
  \!\!\right), \quad
\X=\left(\!\!
  \begin{array}{cc}
    0&i\\
  0&0
  \end{array}
  \!\!\right), \quad
  \Y=\left(\!\!
  \begin{array}{cc}
    0&0\\
  i&0
  \end{array}
  \!\!\right).
  \label{sl2}
  \eeq
We find then
\beq
\fl
I_\th= \frac{1}{\sqrt{\ch(\th)}}
\exp\left(-\tanh(\th)\frac{z^2}{2}\right)
\exp\left(-\ln\ch(\th)z\dz\right)
\exp\left(\tanh(\th)\frac{1}{2}\ddz\right),
\label{Ibarg}
\eeq
and
\beq
\fl
T_\th= \frac{1}{\sqrt{\ch(2\th)}}
\exp\left(\frac{2}{\om}\frac{z^2}{2}\right)
\exp\left(\left[i\frac{\pi}{2}-\ln\ch(2\th)\right]z\dz\right)
\exp\left(\frac{2}{\om}\frac{1}{2}\ddz\right).
\label{Tbarg}
\eeq
To compute the image of $\vp_1(z)$ under $T_\th$, we write
\beq
\vp_1(z) =\exp\left(\frac{\g}{2}z^2\right)f(z) + g(z),
\label{decomp}
\eeq
such that $f(z)\ra c_0\neq0$ and  $g(z)e^{-(\g/2)z^2}\ra 0$ for $|z|\ra\infty$ in $S_{0,2}$. The asymptotic behavior of the image
$T_\th[\vp_1](z)$ is determined by the diverging factor in (\ref{decomp}), which can be expressed through an element of $SL_2(\Cc)$ acting on the constant function,
\beq
\exp\left(\g\frac{z^2}{2}\right)=\exp(\g\X)[1](z).
\eeq
Using now the two-dimensional representation of $T_\th$ and $\X$,
\beq
T_\th=e^{-i\pi/4}
\left(\!\!
\begin{array}{cc}
  i\ch(2\th) & -\sh(2\th)\\
  -\sh(2\th) & -i\ch(2\th)
\end{array}
\!\!\right), \quad
\exp(\g\X)=
\left(\!\!
\begin{array}{cc}
  1 & i\g\\
  0 & 1
\end{array}
\!\!\right),
\eeq
we obtain
\beq
T_\th\exp(\g\X)=e^{-i\pi/4}\exp(\eta_x\X)\exp(\eta_z\Z)\exp(\eta_y\Y),
\eeq
with
\beq
\eta_x(\g)=-\frac{\g+\tanh(2\th)}{\tanh(2\th)\g+1}
=\frac{\frac{2}{\om}-\g}{1-\frac{2}{\om}\g}.
\label{eta}
\eeq
It follows
\beq
T_\th\exp\left(\g\frac{z^2}{2}\right)=
e^{-i\pi/4}\exp\left(\eta_x(\g)\frac{z^2}{2}\right).
\label{image}
\eeq
Solutions of (\ref{eq3}) are mapped by $T_\th$ onto solutions of
\beq
\fl
\vp_2^{(3)}+\left(b_1z+\frac{b_2-2}{z}\right)\vp_2^{(2)}
+(z^2+b_3)\vp_2^{(1)} +\left(\left[2-\frac{E_1}{\om_1}\right]z
+\frac{b_4}{z}\right)\vp_2=0,
\label{eq4}
\eeq
which is obtained from (\ref{eq1}),(\ref{eq2}) after elimination of $\vp_1$.
It has the same singular points as (\ref{eq3}), the exponents of first kind at $z=0$ are $0$, $1$ and $E_1/\om_1+2$. As usual, the non-integer exponent differs from the one for $\vp_1$ by an integer, so that these solutions for $\vp_1$ and $\vp_2$ fulfill together the system (\ref{eq1}),(\ref{eq2}). Clearly, the exponents of second kind at $z=\infty$ are the same as in (\ref{eq3}), with $\r=E_1/\om_1-2$. Therefore, the branching structure of the third solution of (\ref{eq4}) with $\g=0$ has the form shown in Fig.\ref{branch} for the associated solution of (\ref{eq3}).

Lets assume now that the exponent of $\p_1(z)$ in (\ref{decomp}) is $\g=2/\om$.
The exponent of its image under $T_\th$ given by (\ref{eta}) is zero which means that the dominant part of $\p_1(z)$ is mapped onto the recessive part $g(z)$ of $\vp_2(z)$ in the decomposition (\ref{decomp}). Because $T_\th$ is an isometry, normalizable solutions are mapped onto normalizable ones, so the part diverging as $\exp(z^2/\om)$ is mapped onto a less diverging part of a normalizable solution of (\ref{eq4}) which must be analytic at $z=0$. On the other hand, if the exponent of $\vp_1(z)$ is $\g=\om/2$, the exponent of the image $T_\th[\vp_1](z)$ would be infinite according to (\ref{eta}). This indicates that the corresponding solution of (\ref{eq4}) cannot be entire, otherwise it would be described by either $\g=\om/2$ or $\g=2/\om$ in sectors $S_{0,2}$. The only possibility is that non-normalizable solutions of (\ref{eq3}) which are analytic at $z=0$ are mapped by $T_\th$ onto solutions of (\ref{eq4}) which are not analytic at $z=0$, with non-integer exponent $\rho=E_1/\om_1+2$. Recall that those solutions possess an essential but non-isolated singularity at $z=\infty$, because the asymptotic expansion (\ref{regasym}) is valid in the Stokes sector $S^\ast$. Their behavior at infinity cannot be captured by an entire function of order two, indicated by the diverging $\eta_x(\g)$. They are not elements of $\B$ already due to their non-analyticity at $z=0$.

It follows that $T_\th$ maps the set of normalizable solutions of (\ref{eq3}) onto the corresponding set of solutions of (\ref{eq4}). These solutions are all analytic at $z=0$ and have exponent $\g=2/\om$ in sectors $S_{0,2}$. The non-normalizable solutions which are analytic at $z=0$ but have exponent $\g=\om/2$ are mapped onto solutions which are \emph{not} analytic at $z=0$ but have there a branch-cut with exponent $E_1/\om_1+2$. We can now follow the same strategy as for the linear QRM and construct a pair of even functions $\vp_1(z)$, $\vp_2(z)$  satisfying the system (\ref{eq1}),(\ref{eq2}) and both analytic at $z=0$.
If $\vp_2(z)$ and $\vp_1(z)$ have the expansions
\beq
\vp_2(z)=\sum_{m=0}^\infty a_{2m}z^{2m}, \quad
\vp_1(z)=\sum_{m=0}^\infty \ba_{2m}z^{2m},
\label{expans}
\eeq
it follows from (\ref{eq1}) that the coefficients are related by
\beq
\ba_n=\frac{\D}{E_1-n\om_1}a_n,
\label{p1p2}
\eeq
if $E_1/\om_1$ is not an integer. The coefficients of $\vp_1$ are uniquely fixed in terms of the coefficients of $\vp_2$ if both are analytic at $z=0$ without invoking the symmetry. From (\ref{eq2}) one deduces the three-term recurrence relation for the $a_n$,
\beq
\fl
a_{n+2}=\frac{1}{(n+2)(n+1)\G}\left(\left[\frac{\D^2}{E_1-n\om_1}-(n\om_2+E_2)\right]a_n-\G a_{n-2}\right).
\label{p2-recur}
\eeq
The initial conditions can be chosen as $a_{-2}=0$, $a_0=1$. A Poincar\'e analysis of (\ref{p2-recur}) shows that the convergence radius of the expansions in (\ref{expans}) is infinite, as it should be. Both $\vp_1$ and $\vp_2$ are entire and their asymptotics in $S_{0,2}$ have either the exponent $\g=2/\om$ if  they are normalizable or the exponent $\om/2$ if not. The symmetry imposes now the additional relation between $\vp_1$ and $\vp_2$,
\beq
\vp_2(z) = T_\th[\vp_1](z)
\label{symmetry}
\eeq
for all $z\in\Cc$. If $\p_1(z)$ has asymptotics with exponent $\g=\om/2$ in $S_{0,2}$, rendering it non-normalizable, it will be mapped onto a function which is not analytic at $z=0$ and (\ref{symmetry}) cannot be satisfied, because $\vp_2(z)$ is analytic by construction. It follows that the solutions of (\ref{p1p2}) and (\ref{p2-recur}) cannot satisfy (\ref{symmetry}) for generic values of $E$ because they both have asymptotics with $\g=\om/2$. If $E$ is element of the regular spectrum of $\Hp$ and corresponds thus to a certain eigenvalue of the $\Zz_4$-symmetry, condition (\ref{symmetry}) will be satisfied because the image of $\vp_1$ under $T_\th$ is a normalizable solution of (\ref{eq4}) which must be analytic at $z=0$. Therefore, $E$ is an element of the spectrum with eigenstate in $\H_{++}$ if and only if $\vp_{1,2}$ as determined by (\ref{p1p2}),(\ref{p2-recur}) satisfy (\ref{symmetry}). This argument is independent from the presence of the Stokes phenomenon because only sectors $S_{0,2}$ determine the normalizability as $\om/2>0$ and $2/\om >0$. The asymptotics in sectors $S_{1,3}$ play no role, in contrast to the original situation described by (\ref{orig4}) where the exponents of second kind take positive and negative values.

To derive the $G$-function explicitly, we must compute the action of $T_\th$ given in (\ref{Tbarg}) on the series expansion of $\vp_1(z)$ in (\ref{expans}), i.e. on even powers of $z$,
\beq
\fl
T_\th[z^{2m}](z)=\frac{1}{\sqrt{\ch(2\th)}}
\exp\left(\frac{2}{\om}\frac{z^2}{2}\right)
\exp\left(\left[i\frac{\pi}{2}-\ln\ch(2\th)\right]z\dz\right)
P^m_{\om}(z),
\label{t-zm}
\eeq
where $P^m_\om(z)$ is a polynomial in $z$ of degree $2m$,
\beq
P_{\om}^m(z)=\sum_{l=0}^m\om^{l-m}\frac{(2m)!}{(m-l)!(2l)!}z^{2l}.
\eeq
It follows
\beq
T_\th[z^{2m}](z)=\frac{1}{\sqrt{\ch(2\th)}}e^{z^2/\om}
P_{\om}^m\left(\frac{e^{i\pi/2}z}{\ch(2\th)}\right).
\label{t-zm2}
\eeq
It is evident that  expression (\ref{t-zm2}) becomes most simple at $z=0$,
\beq
T_\th[z^{2m}](0)=\frac{1}{\sqrt{\ch(2\th)}}\frac{(2m)!}{\om^m m!},
\eeq
and condition (\ref{symmetry}) would read
\beq
\vp_2(0) = 1 =\frac{(\om^2-4)^{1/4}}{\sqrt{\om}} \sum_{m=0}^\infty \frac{a_{2m}\D(2m)!}{(E_1-2m\om_1)\om^m m!},
\label{first}
\eeq
where the $a_{2m}$ are recursively determined by (\ref{p2-recur}).
However, a Poincar\'e analysis of the series
\beq
f(w) = \sum_{m=0}^\infty \frac{a_{2m}(2m)!}{(E_1-2m\om_1)\om^m m!} w^m
\label{w-series}
\eeq
shows that the convergence radius $R(f)=1$, meaning that $f(w)$ is not defined at $w=1$, which is supposed to yield  $T_\th[\vp_1](0)$. This is easy to understand because the generic form of $T_\th[\vp_1](z)$ has a branch-cut at $z=0$ and each series expansion of it diverges there. The analysis of (\ref{w-series}) gives thus another independent proof of the fact that a non-normalizable but entire solution of (\ref{eq3}) is mapped onto a solution of (\ref{eq4}) which is not analytic at $z=0$.  

\begin{figure}
  \includegraphics[width=0.6\linewidth]{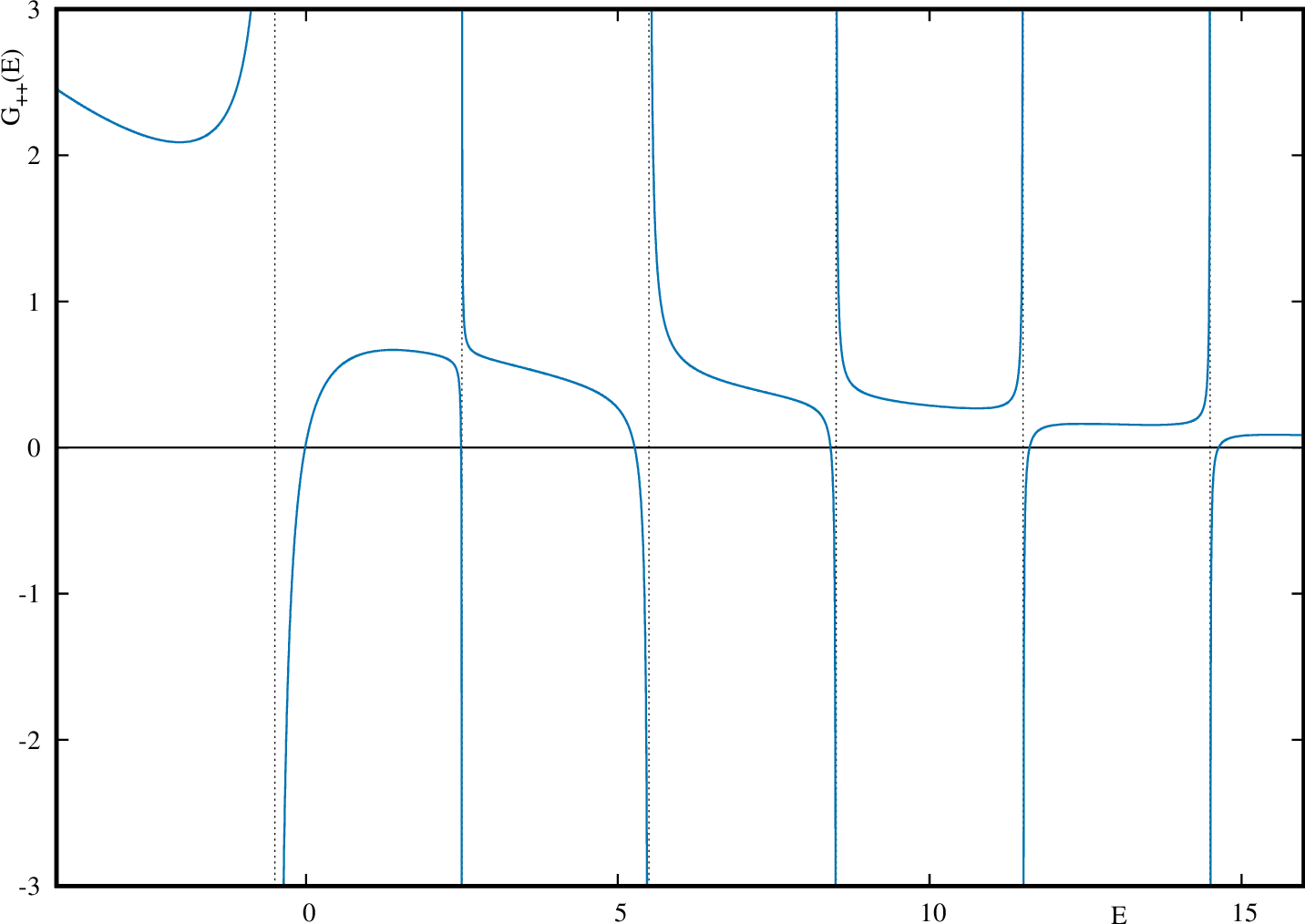}
  \caption{Chen's $G$-function $\Gc(E)$ for the same parameters as in Figs.\ref{zgfunction} and \ref{tgfunction}. This function is characterized by poles at even integer values of $E_1/\om_1$. The regular zeros are located between the poles which determine the average distance between adjacent zeros.}
  \label{cgfunction}
\end{figure}

To derive a $G$-function whose series expansion does converge, we apply $I_\th^{-1}$ to both sides of (\ref{symmetry}),
\beq
\p_2(z)=I^{-1}_\th[\vp_2](z)=T[I_\th^{-1}[\vp_1]](z)=I_\th^{-1}[\vp_1](iz).
\label{Gc1}
\eeq
The expressions in (\ref{Gc1}) are well defined at $z=0$, so it is possible to evaluate the $G$-function at the invariant point $z_0=0$ of the map $T$. This has the advantage that $\p_2(z_0)-\p_1(iz_0)=0$ is sufficient to determine the regular spectrum in $\H_{++}$. If $z_0\neq 0$, one would have two independent conditions $\p_2(z_0)=\p_1(iz_0)$ and $\p_2(-iz_0)=\p_1(z_0)$ \cite{braak_13}.   
Using (\ref{Ibarg}), we find for $I_\th^{-1}[\vp_2](0)$,
\beq
I_\th^{-1}[\vp_2](0) = \frac{1}{\sqrt{\ch(\th)}}\sum_{m=0}^\infty a_{2m}\frac{\tanh(|\th|)^m(2m)!}{2^m m!}.
\label{p2zero}
\eeq
The Poincar\'e analysis of the series
\beq
\tilde{f}(w)=\sum_{m=0}^\infty a_{2m}\frac{\tanh(|\th|)^m(2m)!}{2^m m!} w^m
\eeq
yields the convergence radius $R(\tilde{f})=1+\sqrt{\om^2-4}/\om>1$, therefore the expression (\ref{p2zero}) is given by an absolutely converging series. {\it A fortiori} the same applies to $I^{-1}_\th[\vp_1](0)$ and we can define a $G$-function as $\Gc(E)=\p_2(0)-\p_1(0)$, that is
\beq
\Gc(E)=\sum_{m=0}^\infty\left(1-\frac{\D}{E_1-2m\om_1}\right)a_{2m}\frac{\tanh(|\th|)^m(2m)!}{2^m m!}.
\label{Gc}
\eeq
After adjusting notation, we find $\Gc(E)=G^{1/4}_+(x(E))$ which is the $G$-function for sector $\H_{++}$ given in (16) of \cite{duan_16} and has been originally proposed ten years ago by Chen {\it et al.} in \cite{chen_12}.
By examination of the normalizability properties of formal solutions to the Schr\"odinger equation in the Bargmann space, we have demonstrated that Chen's $G$-function yields the complete regular spectrum in $\H_{++}$ (the other sectors require minor modifications of the derivation). It is shown in Figure~\ref{cgfunction}. 

The major qualitative difference of this $G$-function compared with the others proposed by Zhang \cite{zhang_17}, Trav{\v e}nec \cite{travenec_12} and Maciejewski/Stachowiak \cite{mac_17} is the presence of simple poles at even integer values of $E_1/\om_1$, that means at energies
\beq
E^{(m)}=\left(2m+\frac{1}{2}\right)\sqrt{\om^2-4}-\frac{\om}{2}, \qquad m\in\Zpl_0.
\label{poles}
\eeq
For most values of the parameters $\om$ and $\D$ all eigenenergies are regular and located between the poles given by (\ref{poles}). According to the conjecture formulated in \cite{braak_11}, the number of zeros between adjacent poles is restricted to be 0, 1 or 2. This conjecture has been numerically verified in the linear and 2pQRM and can be derived analytically for small $\D$. It plays here a similar role as the string hypothesis in systems solvable by the Bethe ansatz \cite{eckle_19}. As discussed in \cite{duan_16}, the conjecture entails that the average distance between eigenenergies in $\H_{++}$ is given by the distance between the poles, $2\sqrt{\om^2-4}$. The poles are equally spaced along  the whole real axis for $E \ge E^{(0)}=(1/2)(\sqrt{\om^2-4}-\om)$ and become more dense as $\om$ approaches the critical value $\om_c=2$ from above -- the distance between adjacent energy levels goes to zero in this limit. At $\om_c$, the real axis above the threshold value $E_0=-1$ becomes densely covered with levels, indicating the transition to a continuous spectrum as derived in  section \ref{general} for the critical point itself. In this way the qualitative properties of the spectrum as the system approaches $\om_c$  can be deduced from the known pole structure of $\Gc(E)$. There is no ``collapse'' of the spectrum at the point $\om_c$ to the single energy $E_0$ as falsely inferred from exact diagonalization in a truncated state space but instead the transition to a continuous spectrum at $\om_c$, spanning the whole real axis from $E_0$ to infinity. The exceptional spectrum in $\H_{++}$ consists of those energy levels which coincide with a pole energy $E^{(m)}$ in full analogy with the linear QRM. In this case the pole in $\Gc(E)$ is lifted by a zero of the expansion coefficients in  (\ref{Gc}) and $\Gc(E^{(m)})$ is finite. It is easy to see that those eigenstates are exactly the special solutions found in \cite{emary_02}. They appear if the asymptotic expansion (\ref{regasym}) of the third solution to (\ref{eq3}) is convergent because the recurrence (\ref{recur-c}) breaks off after $m$ steps, leading to another normalizable entire solution, a polynomial. The exceptional level is thus doubly degenerate. We have seen theat the presence of poles in $\Gc(E)$ is instrumental to determine the approach to the collapse point without numerical evaluation of the $g$-function. On the other hand, it may seem detrimental to have access only to the regular part of the spectrum. As shown in \cite{li_add_16,kimoto_21}, it is possible to obtain a $G$-function for the linear QRM whose zeros give the complete spectrum including the degenerate and non-degenerate exceptional parts by multiplying $G_\pm(E)$ with appropriate Gamma-functions. The same is possible for $\Gc(E)$. This may be of importance to make connections with the spectral zeta-functions associated with $H_R$ and $\Hp$ \cite{wakayama_17,kimoto_21}.

The degeneracies of Juddian type, between states in $\H_{++}$ and $H_{+-}$, are connected to the poles of $\Gc$ and $G_{\scriptscriptstyle{+-}}$, where $E_1/\om_1$ is a positive even integer. Likewise the degeneracies between $\H_{-+}$ and $H_{--}$ correspond to lifting of a pole in $G_{\scriptscriptstyle{-\pm}}$. The poles correspond to $E_1/\om_1$ being a positive odd integer. The degeneracies appearing between even and odd eigenstates with $E_1/\om_1$ a half-integer are not reflected in these $G$-functions. They correspond to a factorization property of (\ref{orig4}) for special values of the parameters $\om$, $\D$ which does not follow from the $\Zz_4$-symmetry alone \cite{mac_19}.

\section{Conclusions}
\label{concl}
Many qualitative aspects of the spectrum of the two-photon quantum Rabi model which define the three different parameter regimes, $\om>2$, $\om=2$ and $\om<2$, can be inferred without numerical computation by examining the singularity structure of the Schr\"odinger operator in the Bargmann space of entire functions. In these regimes, the spectrum is either pure point ($\om>2$) or comprises a continuous and a discrete part ($\om=2$). For $\om<2$ the spectrum is at least partly continuous with generalized eigenstates resembling asymptotically those of the inverted harmonic oscillator. At the ``collapse point'' $\om=2$, the effective potential following from (\ref{phi1eq}), (\ref{phi2eq}) depends on the energy eigenvalue so that no direct mapping to a one-dimensional problem without spin degree of freedom is possible. Nevertheless qualitative aspects of the spectrum may be derived from the analogy (section \ref{sec-critical}).

The various generalized spectral determinants proposed in the regime $\om>2$ have different qualitative properties depending on their derivation. While Zhang's $G$-function \cite{zhang_17}, based on a continued fraction, and Chen's $G$-function \cite{chen_12} based on a Bogoliubov (squeezing) transformation, both possess poles on the real axis located between the zeros (eigenvalues of $\Hp$), their position is known analytically only in the latter case. Neither Trav{\v e}nec's \cite{travenec_12} nor Maciejewski's $G$-functions \cite{mac_17} exhibit poles or other known qualitative features, like the position of maxima or minima, to infer the distribution of the zeros without explicit numerical computation. The average distance of energy levels, while independent from the coupling in the linear QRM, changes dramatically in the 2pQRM if the critical coupling $g=\om/2$ is approached from below (or $\om \ra 2$ from above for fixed $g=1$). The levels coalesce on the real axis above the threshold energy $E_0=-1$ to form the continuous part of the spectrum at the critical point $\om=2$. Because the average distance of levels is given by the pole distance in Chen's $G$-function, $\D E=2\sqrt{\om^2-4}$, the exponents of this critical behavior can be computed exactly \cite{duan_16}. Moreover, Chen's $G$-function yields the quasi-exact Juddian points of the spectrum  in the same way as the corresponding $G$-function of the linear QRM. While the validity of this $G$-function has been confirmed numerically in great detail, a rigorous derivation based on the normalizability of the wave functions was missing. As the numerical checks fail close to the critical point due to  truncation of the Hilbert space, the rigorous proof of validity provided here seems not only of formal but also of practical importance.

\medskip
\textbf{Acknowledgements} \par 
 I wish to thank F. Hiroshima for discussion of the case $\om<2$. This work was funded by the Deutsche Forschungsgemeinschaft (DFG, German Research Foundation) under grant no.: 439943572. 
 \par
 \vspace{10mm}
 \medskip
 \textbf{References}\\
%
%

\end{document}